\documentclass[12pt] {iopart}
\usepackage{amssymb}
\usepackage{graphicx}
\begin{document}
\title{ Transverse even effect and Hall effect in YBaCuO superconductor}
\author{I Jane\v{c}ek\dag and P Va\v{s}ek\ddag}
\address{\dag\ University of Ostrava, Dvorakova 7, 70103 Ostrava, Czech Republic}
\address{\ddag Institute of Physics ASCR, Cukrovarnick\'a 10,\\ 162 53 Praha
6, Czech Republic\\}
\ead{vasek@fzu.cz}



\begin{abstract}
The Hall resistance was measured by the van der Pauw method on $YBa_2Cu_3O_x$ samples in magnetic fields up to 5 T.
In region from 0 to 0.2 T the part of the resistance due to transversal electric field, which is even in magnetic field  has been observed just below the critical temperature. This part is gradually suppressed in higher field. Only the Hall resistance (due to odd transversal voltage) has been   observed in magnetic field 5 T. The Hall resistance has a typical temperature dependence with negative minimum  below the critical temperature. The anisotropy of the  Hall resistivity has been observed in this region. We have also tested the reciprocity theorem which is not valid near  critical temperature.     
\end {abstract}
\submitto{\SUST}

\maketitle

\section{Introduction}

Two effects can be observed in high-temperature superconductors due to the existence of two different (with respect to the reversal of  magnetic field) transversal electric fields. The Hall effect is a result of the electric field, which is transversal to transport current when  magnetic field is applied. This electric field is an odd function of the direction of the magnetic field.  In terms of the resistivity tensor this is its antisymmetric part. However,  transversal electric field connected with  symmetric part of the resistivity tensor can also be  detected. This  field is an even function of the magnetic field. The existence of such even transversal field is called the transverse even effect or simply the even effect. It is observed in  high temperature superconductors near critical temperature (namely in the low field region) (\cite{Java2}and references therein). Moreover, the transversal electric field can be detected also in zero magnetic field \cite{Fra,Va,Java,Yamo}.
Both  effects are followed by the violation of the reciprocity theorem \cite{Java2,Java}, which says that in an four-point measurement an interchange of the voltage and current contacts does not change value of the voltage detected \cite{But1,But2,But3}. In a magnetic field the direction of the magnetic induction must also be reverted (the magnetic field form of the reciprocity theorem). 
 
In this article we present results obtained in magnetic field for the same sample of YBaCuO that we measured before in zero magnetic field {\cite{Java}.

\section{Experiment}

We have measured four-terminal resistances in magnetic fields up to 5 T. For the detection of the transverse resistivity the van der Pauw method {\cite{Pauw1,Pauw2,Pauw3} has been used. The sample was in the form of a thin slab obtained by grinding  the highly  oriented pellets of the YBaCuO material prepared by seeded melt-texturing process. The c-axis was perpendicular to the slab surface. The  critical temperature $T_c$ was 89.5 K. 
Three single resistances $R_{AB,CD}$, $R_{BC,DA}$, and $R_{AC,BD}$ (first two letters describe current contacts, last two of them determine voltage contacts) have been measured. Resistance due to transversal electric field can be calculated using  three-resistance formula
\begin {equation}
R^{\mathrm{(III)}}_{ABCD}\equiv -R_{AB,CD}+R_{BC,DA}+R_{AC,BD}.
\end {equation}
Moreover,  three resistances for cyclic permutations of the contact have been measured as well. Thus we have also  $R_{DA,BC}$, $R_{AB,CD}$, and $R_{DB,AC}$ and corresponding formula is
\begin {equation}
R^{\mathrm{(III)}}_{DABC}\equiv -R_{DA,BC}+R_{AB,CD}+R_{DB,AC}.
\end {equation}
Because these three-resistance combinations incorporate contribution from even and odd transversal electric fields, the same resistances have been measured  also for opposite magnetic field direction to separate odd and even part of $R^{\mathrm{(III)}}$ resistances. The odd part is equal to the Hall resistance $R^H$.

The temperature dependence of the described  resistance combinations has been determined  from the single resistances measured.  Temperature intervals of the measurements always included sufficiently wide surroundings of the superconducting transition. Corresponding temperature dependence for the odd and even parts of the $R^{\mathrm{(III)}}$ resistances are presented in Fig.1 (a-e).    Moreover, we have measured dependence of these quantities on the magnetic field for two temperatures selected near the critical temperature. See Fig.2 (a-b).

From Fig.1 and Fig.2 one can see that zero field effect continues as even effect in weak magnetic field. Here the Hall effect is small. For higher magnetic field (above 0.2 T) the even effect gradually vanishes and for high field (over 0.5 T) one can observe only the Hall effect. The Hall resistance has a typical temperature dependence with the change of sign and  a minimum  below  critical temperature. The depth of the minimum decreases with increasing  magnetic field in contrast  to the Hall resistance value in normal state which increases.
\vspace*{-1cm}
\begin{figure}[b]
\begin{center}
\leavevmode
\hspace*{1cm}
\includegraphics[width=12.0cm]{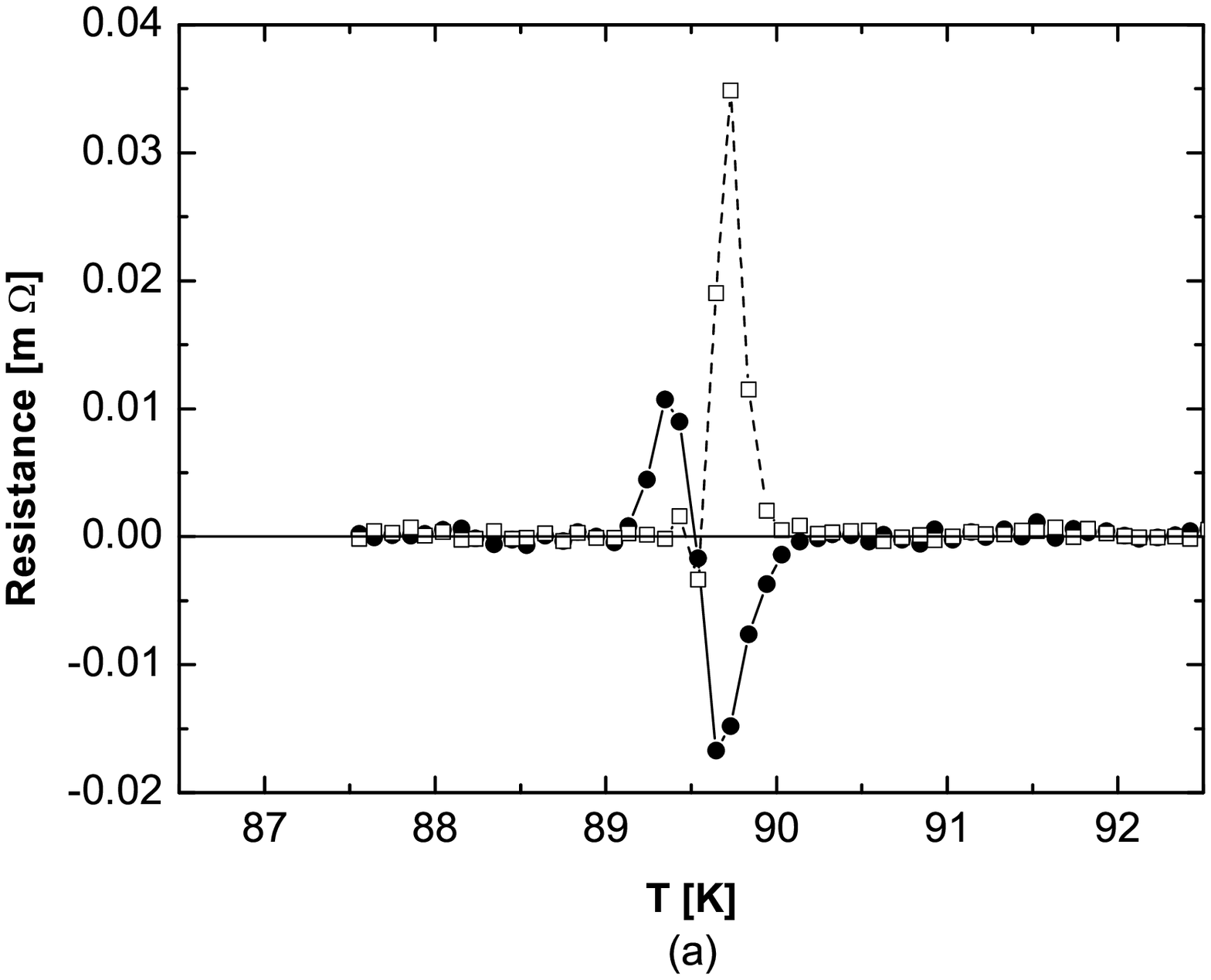}
\vspace*{-1.5cm}
\end{center}
\end{figure}
\vspace*{1.5cm}

\section{Discussion}

Recently  we have  suggested  a model  explaining the existence of the  transversal electric fields, which is observed in zero magnetic field  and/or in non-zero field in high temperature superconductors \cite{Java2,Java}. This model  extends the  guided motion model based on existence of special pinning force which hustles vortices to move in preferred direction \cite{Kop}. Observations of the above mentioned transversal electric fields are followed also by the reciprocity theorem breaking, which implies the existence of  an additional sample characteristic that does not obey the time reversal symmetry( \cite{Java2,Java}).  One of the candidates for such characteristic is the spontaneous magnetisation in YBaCuO superconductors \cite{Car}. \\
\newpage
\vspace*{-1cm}
\begin{figure}[h]
\begin{center}
\leavevmode
\hspace*{1cm}
\includegraphics[width=12.0cm]{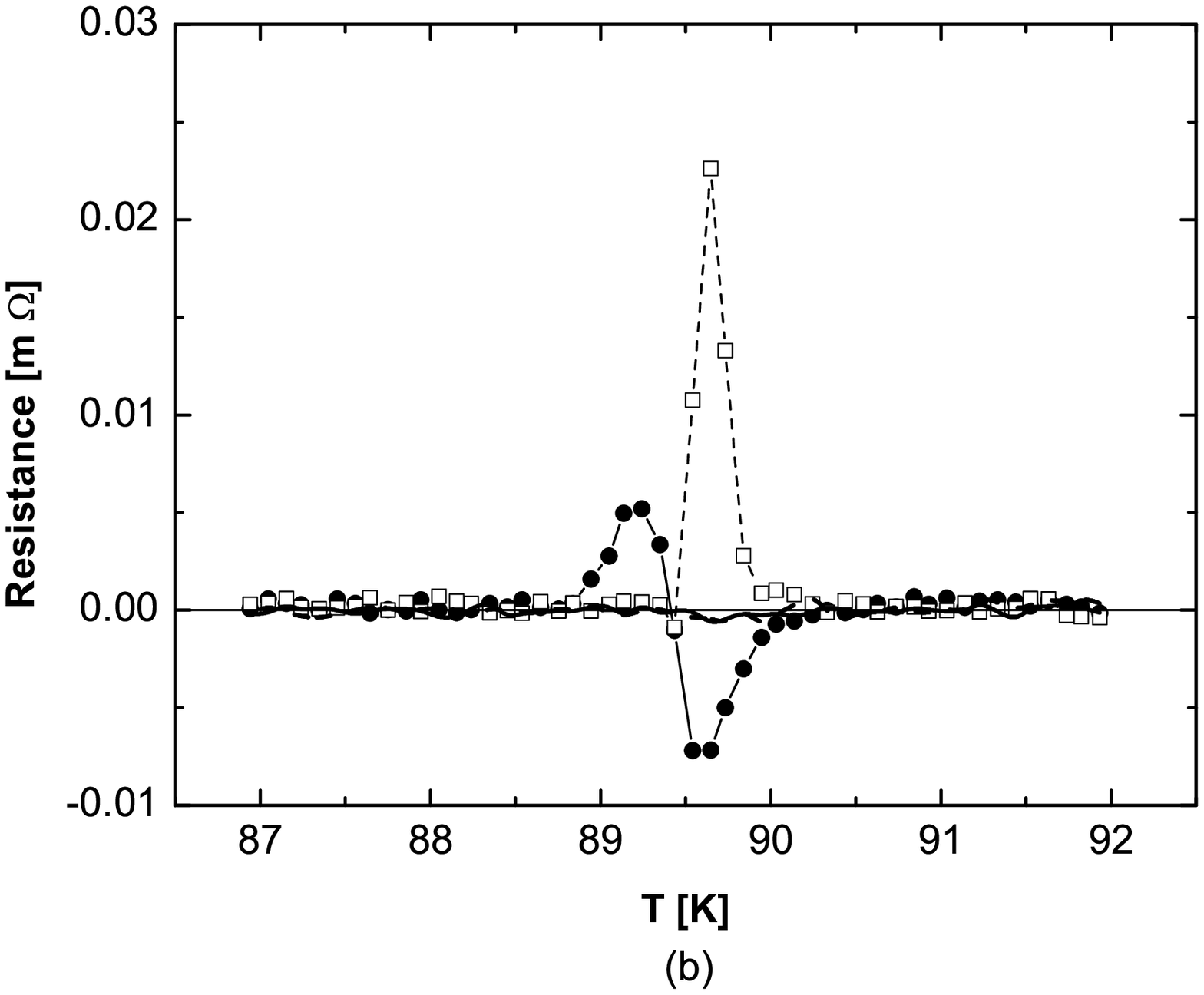}
\vspace*{-1.5cm}
\end{center}
\end{figure}
 
\vspace*{-1cm}
\begin{figure}[h]
\begin{center}
\leavevmode
\hspace*{1cm}
\includegraphics[width=12.0cm]{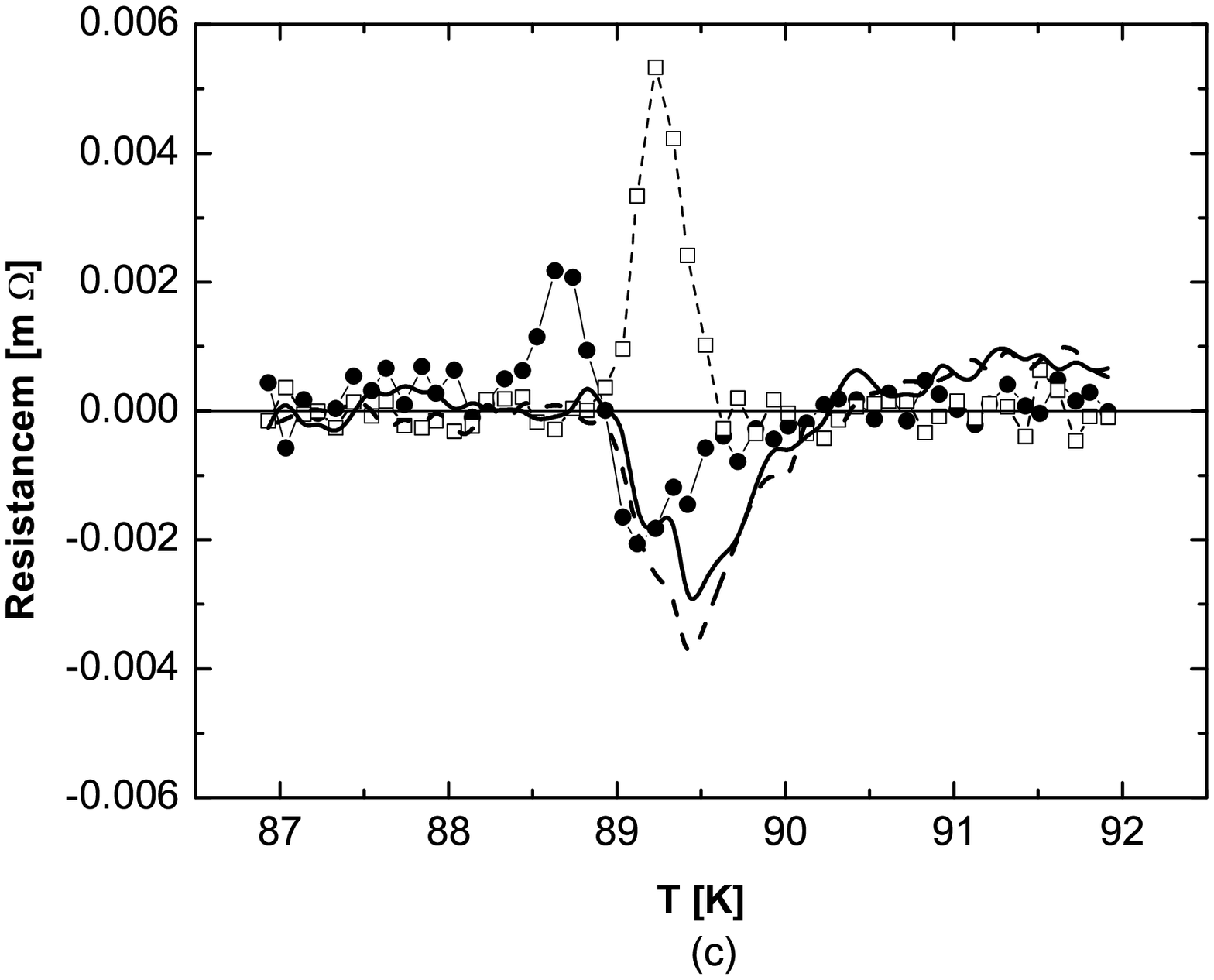}
\vspace*{-1.5cm}
\end{center}
\end{figure}
\newpage
\vspace*{-1cm}
\begin{figure}[h]
\begin{center}
\leavevmode
\hspace*{1cm}
\includegraphics[width=12.0cm]{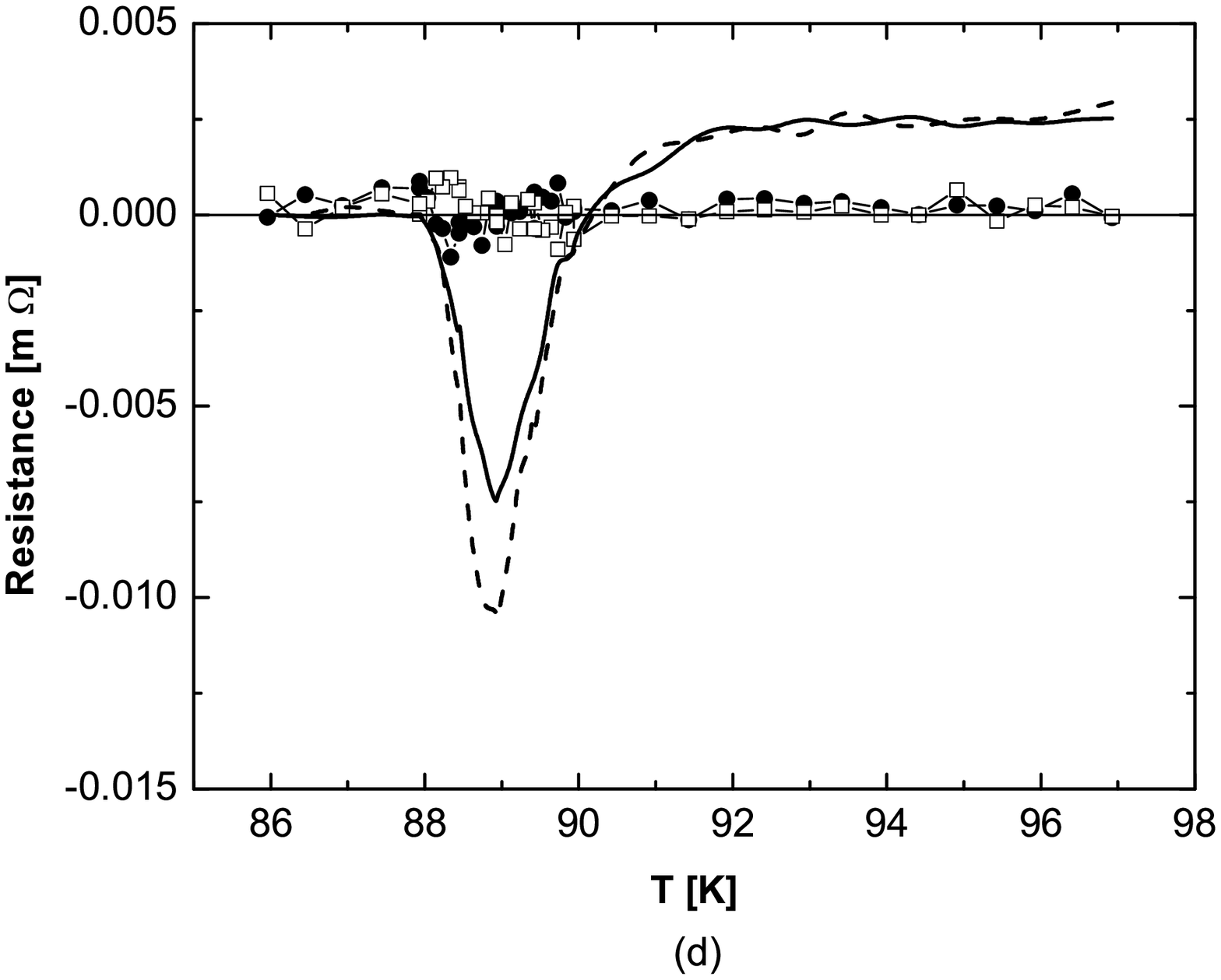}
\vspace*{-1.5cm}
\end{center}
\end{figure}

\vspace*{-1cm}
\begin{figure}[h]
\begin{center}
\leavevmode
\hspace*{1cm}
\includegraphics[width=12.0cm]{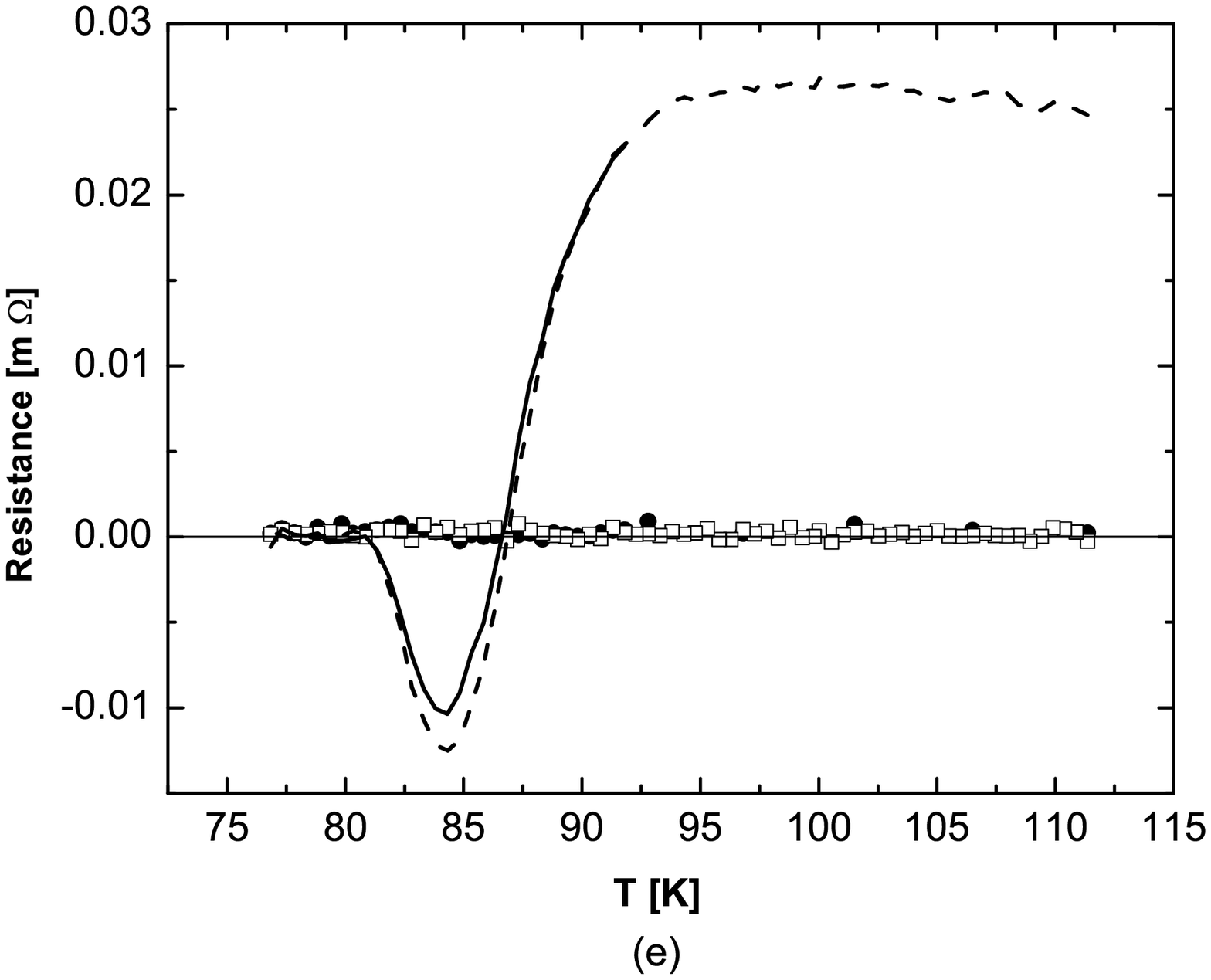}
\caption{Temperature dependence of resistivities:   $ R^{\mathrm{(III)}}_{ABCD}$  (full circle)  and $ R^{\mathrm{(III)}}_{DABC}$  (square)  for $(a)$ $B=0$ T; resistivities  $R^{\mathrm{(III)},{even}}_{ABCD}$  (full circle), $R^{\mathrm{(III)},{even}}_{DABC}$  (square), $R^{\mathrm{(III)},{odd}}_{ABCD}$  (full line), $R^{\mathrm{(III)},{odd}}_{ABCD}$  (dashed line) for  (b) $B=0.05$ T, (c) $ B= 0.2$ T, (d) $B=0.5$ T, (e) $B= 5$ T.}
\label{Fig}
\end{center}
\end{figure}
\newpage
\vspace*{-1cm}
\begin{figure}[h]
\begin{center}
\leavevmode
\hspace*{1cm}
\includegraphics[width=12.0cm]{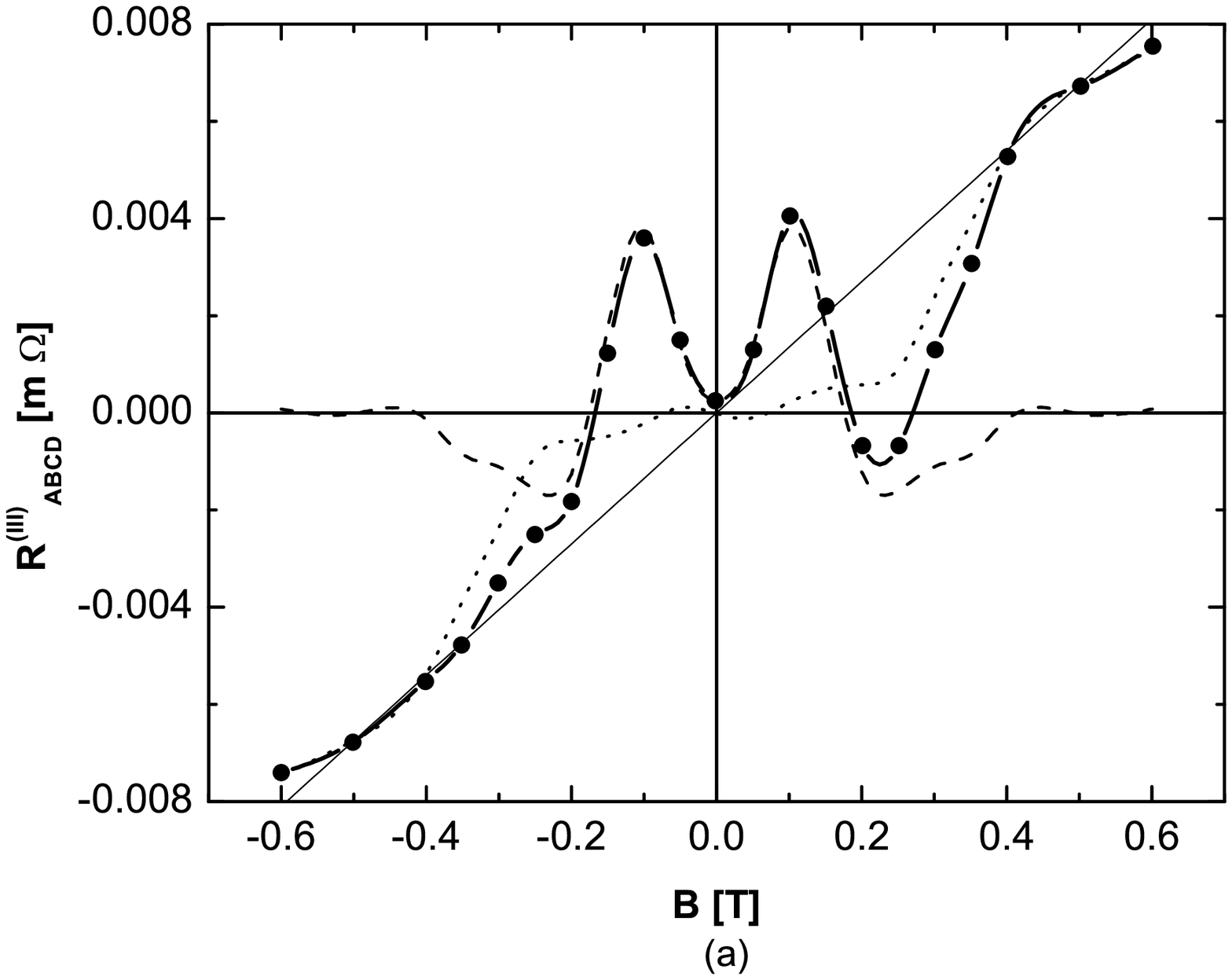}
\vspace*{-1.5cm}
\end{center}
\end{figure}

\vspace*{-1cm}
\begin{figure}[h]
\begin{center}
\leavevmode
\hspace*{1cm}
\includegraphics[width=12.0cm]{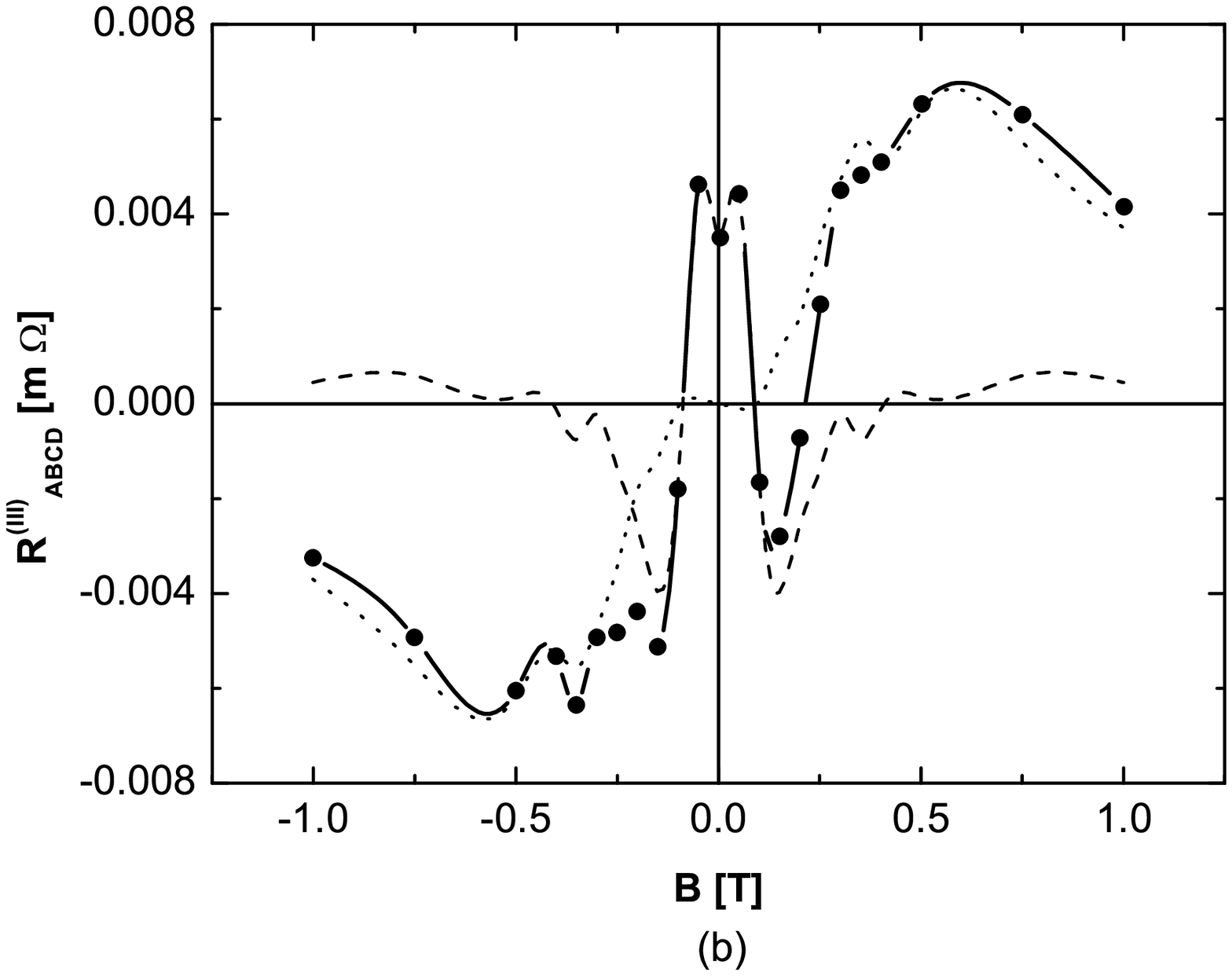}
\caption{ Magnetic field dependence of the resistivities  $R^{\mathrm{(III)}}_{ABCD}$  (full circle), $R^{\mathrm{(III)},{even}}_{DABC}$  (dashed line), $R^{\mathrm{(III)},{even}}_{ABCD}$ (dotted line) for $(a)$ $T= 89$ K and $(b)$ $T= 89.2$ K.} 
\label{Fig}
\end{center}
\end{figure}
\newpage
In the guided motion model the field 0.2 T can be the field, where the Lorentz force exceeded  magnitude of the pinning  forces responsible for guided movement of vortices. If the spontaneous magnetisation model is valid for this YBaCuO material, we suppose that  this field  can be just the field where the inner magnetic moment turns to the direction of the external magnetic field.

We have tried to test the validity of the magnetic field form of the reciprocity theorem in YBaCuO. The way used in zero magnetic field \cite{Java}, where all resistances can be measured together, cannot be used in a non-zero magnetic field.  The  resistances have to be measured in different temperature runs for  opposite magnetic fields. For YBaCuO sample  the single resistances can be  affected in this case by systematic errors due to an temperature shift  \cite{Java} and they cannot thus be used  for the reciprocity theorem test. However, it has been found in the experiment that  three-resistance combinations defined in Eq.1 and Eq.2 are not affected by these effect and the same holds for derivatives of the temperature dependencies of the single resistances. The second fact can be theoretically used to correct for a systematic temperature shift for opposite fields, but in practice this way has difficulties. We use thus  for the magnetic field reciprocity theorem testing the next procedure based on the above mentioned insensibility of the three-resistance combination to the temperature shift.

One can get following equation by adding of  Eq.1 (for $+\vec{B}$) and Eq.2 (for $-\vec{B}$):\\ 

 $A\equiv [R^{\mathrm{(III)}}_{ABCD}(\vec{B}) + R^{\mathrm{(III)}}_{DABC}(-\vec{B})]/2$
\begin {equation}
= -R_{AB,CD}^{(odd)} (\vec{B}) + D_{BCDA}(\vec{B}) +D_{ACBD}(\vec{B}) , \\
\end {equation}
where the first term (with negative sign)\\

$R_{AB,CD}^{(odd)} (\vec{B}) = [R_{AB,CD}(\vec{B}) -R_{AB,CD} (-\vec{B})]/2$\\\\
is odd part of resistance $R_{AB,CD}$.  The second term\\

$D_{BCDA}(\vec{B})=~ [R_{BC,DA}(\vec{B}) - R_{DA,BC}(-\vec{B}) ] / 2 $\\\\
and third term \\

$D_{ACBD}(\vec{B})=~ [R_{AC,BD}(\vec{B}) - R_{BD,AC}(-\vec{B}) ] / 2 $\\\\
represent the  deviation from magnetic field form of the reciprocity theorem for  the cyclic (BC,DA) and the cross (AC,BD) combination of contact, respectively.\\

\vspace*{-1cm}
\begin{figure}[t]
\begin{center}
\leavevmode
\hspace*{1cm}
\includegraphics[width=12.0cm]{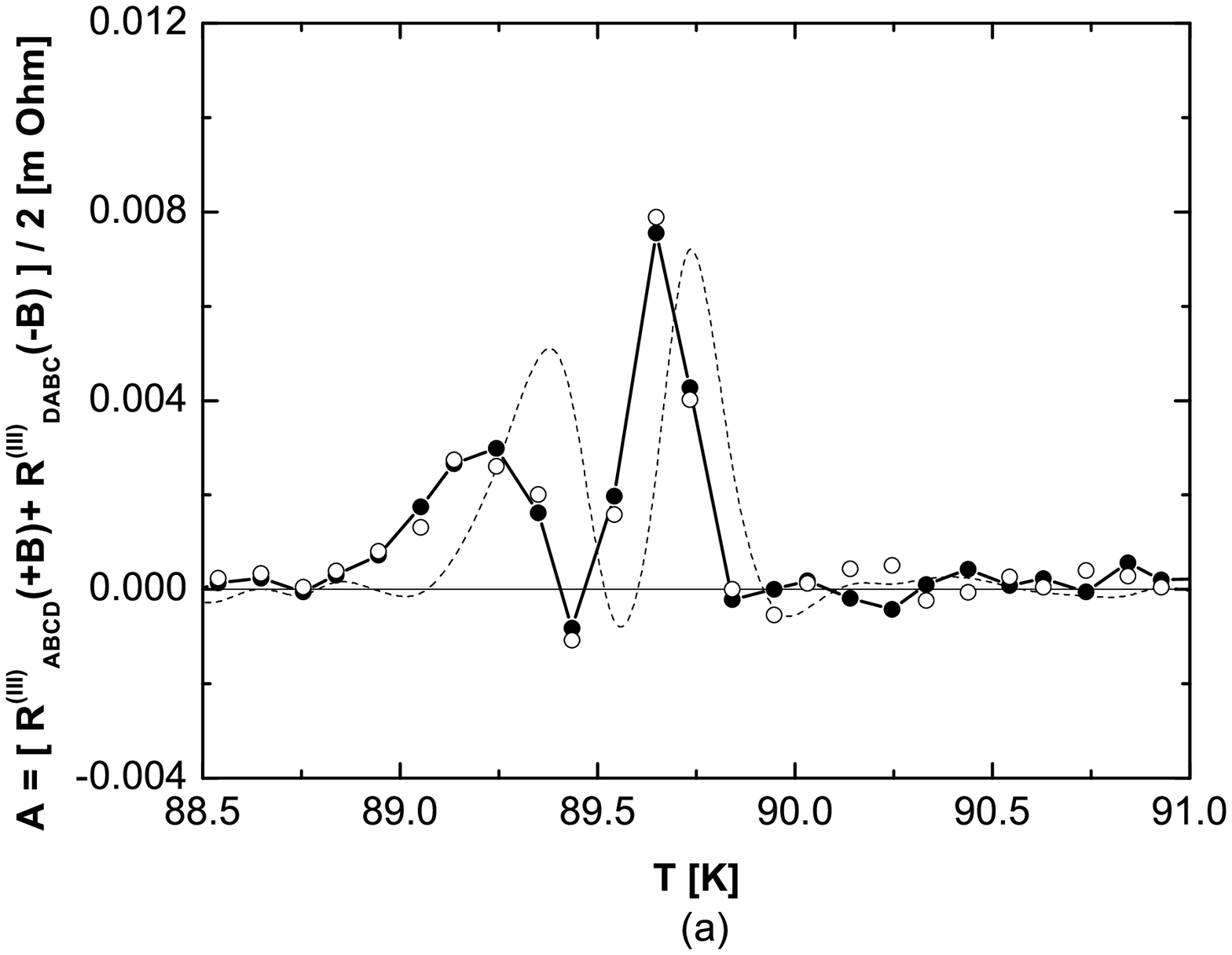}
\vspace*{-1.5cm}
\end{center}
\end{figure}
 
\vspace*{-1cm}
\begin{figure}[h]
\begin{center}
\leavevmode
\hspace*{1cm}
\includegraphics[width=12.0cm]{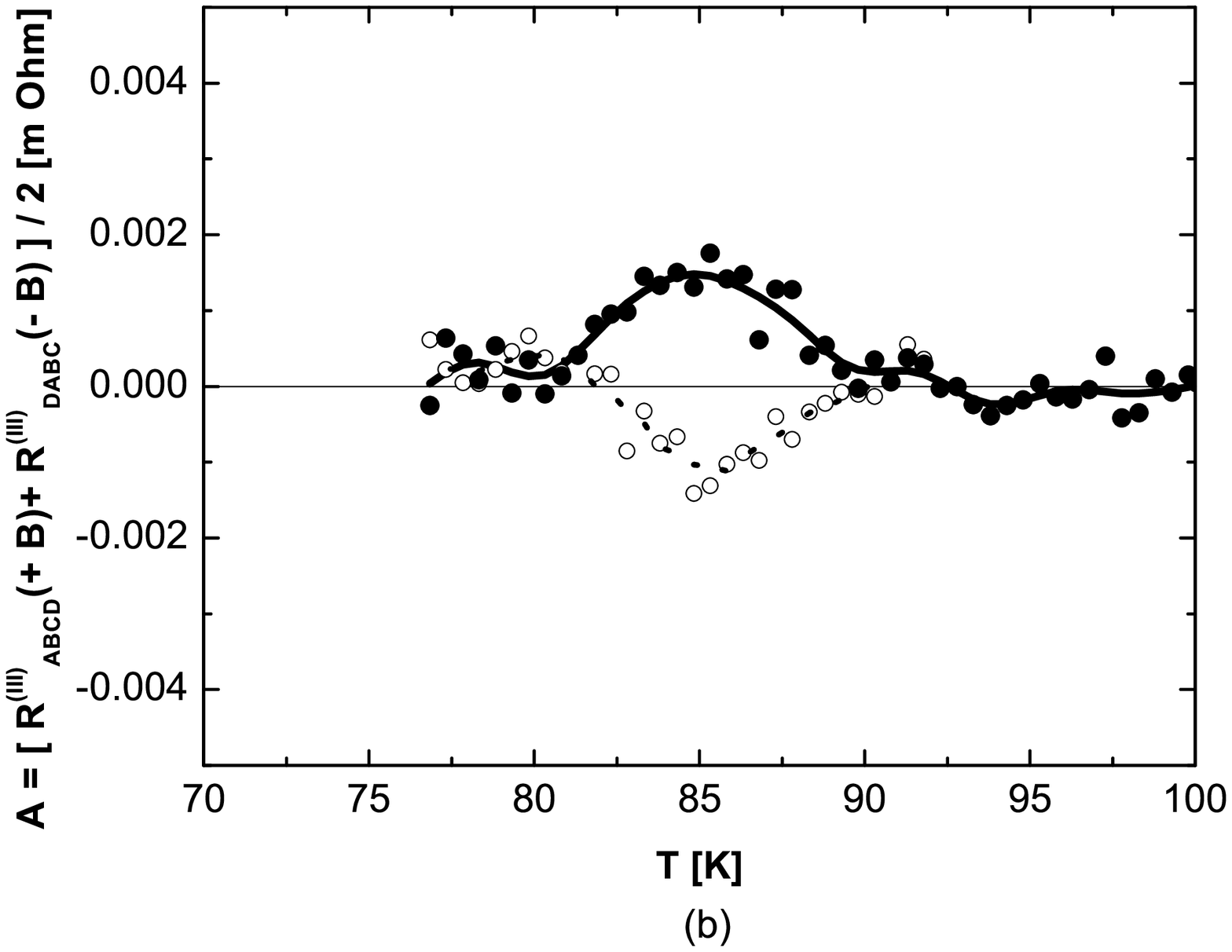}
\caption{Temperature dependence of $A$ for different magnetic fields $(a)$  $B= 0$ T  (dotted line), $B=+0.05$ T  (full circle), $B= -0.05$ T  (open circle); $(b)$ $B= + 5$ T  (full circle), $B= -5 $T (open circle).}
\label{Fig}
\end{center}
\end{figure}
\newpage
First term is zero, if contacts ABCD are exactly on circumference of the sample (a contacts on the same current line can detect only a longitudinal voltage, thus the odd transversal voltage due to the Hall effect cannot be detected).  If the reciprocity theorem is valid the last two terms must be  zero.

The shift of the contacts from circumference of the sample measured is indeed negligible because we observe $A=0$ in the normal state, where the reciprocity theorem is valid and  the Hall voltage is nonzero. Thus a difference $A$ from zero means that magnetic field form of the reciprocity theorem is not valid.

If there is any difference between  temperature dependencies of  $R^{(III)}_{ABCD}(\vec{B})$ and $-R^{(III)}_{DABC}(-\vec{B})$  anywhere, the $A$ will be not zero here. From Fig.1 one can see that the $R^{(III),even}_{ABCD}$ and the $-R^{(III),even}_{DABC}$  and/or  $R^{(III),odd}_{ABCD}$ and the $R^{(III),odd}_{DABC}$  differ only near the critical temperature, i.e.   $A(\vec{B})$ is nonzero around $T_c$.  For  low magnetic field  the dependencies  are similar to those observed in the zero magnetic field and $A(\vec{B})$ is approximately even function of B. For field over 0.2 T we observe also odd part of this deviation. Curves $A(\vec{B})$ for 5 T and 0.05 T are plotted in Fig. 3a and Fig. 3b. Thus the reciprocity theorem is not valid in YBaCuO near the critical temperature.

This observation seems to be in accordance with our simple analysis in \cite{Java2}, hence\\ 

$D_{ACBD}(\vec{B})=~ [R_{AC,BD}(\vec{B}) - R_{BD,AC}(-\vec{B}) ] / 2 =$ \\

$\hspace*{1cm}= [R^{\perp,S}_{AC,BD}(\vec{B}) - R^{\perp,S}_{BD,AC}(-\vec{B}) ] / 2 + [R^{\perp,A}_{AC,BD}(\vec{B}) - R^{\perp,A}_{BD,AC}(-\vec{B})] / 2$ \\

$\hspace*{1cm} = [R_{AC,BD}^{\bot, S}(\vec B) + R_{DB,AC}^{\bot ,S} (\vec B)]/2 + [R_{AC,BD}^{ \bot ,A} (\vec B) - R_{DB,AC}^{\bot ,A} (  \vec B)]/2$\\

Here $R_{kl,mn}^{\perp, S}$ and $R_{kl,mn}^{\perp,A}$ are parts of the resistance $R_{kl,mn}$, connected with transversal resistivity $\rho_{\perp}$(off-diagonal element of the local resistivity tensor). The  $R_{kl,mn}^{\perp, S}$ arises from the  off-diagonal element of the symmetric part of the resistivity tensor, which is even  in magnetic field (the even effect). The resistance $R_{kl,mn}^{\perp,A}$, which comes  from the antisymmetric part,  is due to the Hall effect. The Hall field, which is transversal to current, is odd function of the magnetic field. Thus the deviation consists from two terms,  a sum of the even effect parts for both configurations (ABCD and DABC) and the  difference of the Hall effect parts (odd in magnetic field)  for both configurations. If the Hall resistivity is isotropic ( i.e. $\rho_{xy}^{(odd)} = -\rho_{yx}^{(odd)}$), the second term is zero. From Fig.1d and Fig.1e one can see that the Hall resistance values $R^H_1 \equiv R_{AC,BD}^{ \bot ,A}=R^{(III),odd}_{ABCD}$ and $R^H_2 \equiv R_{DB,AC}^{ \bot ,A}=R^{(III),odd}_{DABC}$ differ in region of the transition (region of the minimum), which is region where we observe odd part of the $A(\vec{B})$.

In BiSrCaCuO materials \cite {Java2}  we do not observe any relevant difference in the Hall resistance obtained from the basic and the cyclically permutated combination. The difference in our YBaCuO material may originate from anisotropy of the Hall resistivity (i.e. $\rho_{xy}^{(odd)} \neq -\rho_{yx}^{(odd)}$).
 
In \cite {Skhlov} the phenomenological model of guided motion of the vortices is presented. There the twin boundaries are considered as guiding pinning canals but this  phenomenological model can be also  used for other guiding mechanisms. The ad hoc postulating of the anisotropic phenomenological coefficients in the steady state equation of vortex motion leads to anisotropic Hall resistivity. The predicted existence of the part of the longitudinal resistivity, which is odd in magnetic field, is therefore a consequence of the presumed anisotropy of the coefficients. However, this fact is supported by observation in \cite{Prod}.
  
It seems to us that considering the basic symmetry this fact cannot be accepted without existence of some additional anisotropy, which is not only a crystallographic anisotropy or anisotropy due to planar defects. The additional anisotropy must be connected with the  existence of a characteristic variable, which changes sign after time reversal but which is independent on external magnetic field. This variable must appear in equation of the motion.

In the next, we analyse the rotational transformation of the resistivity tensor and behaviour of its components under magnetic field inversion to look for the possible origin of the observed effects. 
Local orthogonal system of coordinates is chosen with first axis parallel to local direction of current density $\vec{j} = (j_x, j_y)$ to obtain expression for parallel and orthogonal part of electric field as function of angle $\theta$  between  $\vec{j}$ and direction of x-axis. In common case of the  laboratory orthogonal system (x,y,z) the following transformation relations are valid\\ 

\begin {tabular}{ll}
$E_x = \rho_{xx} j_x + \rho_{xy} j_y$ & $E_{\parallel}=[\rho_{xx} cos^2\theta + \rho_{yy} sin^2\theta  + (\rho_{xy} + \rho_{yx})cos\theta sin\theta] j   $\\
$E_y = \rho_{yx} j_x + \rho_{yy} j_y$ & $E_{\perp}= [\rho_{yx} cos^2\theta - \rho_{xy} sin^2\theta + (\rho_{xx}- \rho_{yy})cos\theta sin\theta] j$
\end {tabular}.\\

In high temperature superconductors the axes of laboratory orthogonal system can be chosen in direction of the crystallographic axes. (The same relations  can be used in case of planar defect, but  axis x or y must be choosen parallel to plane of this defect. See e.g. \cite {Mat}.)  In this special selection of coordinates the symmetrical part of tensor is diagonal and  can be represented only by non-zero diagonal elements  $\rho_{xx}$ and $\rho_{yy}$, which can differ as a consequence of the crystallographic anisotropy or of a planar defects. The antisymmetrical part is represented by off-diagonal element $\rho_{xy} = -\rho_{yx}$. So, the above relations can be simplified. \\

\begin {tabular}{ll}
$E_x = \rho_{xx} j_x - \rho_{yx} j_y$ & $E_{\parallel}=[\rho_{xx} cos^2\theta + \rho_{yy} sin^2\theta] j  $\\
$E_y = \rho_{yx} j_x + \rho_{yy} j_y$ & $E_{\perp}= [\rho_{yx} + (\rho_{xx}- \rho_{yy})cos\theta sin\theta] j$
\end {tabular}\\

Moreover, we suppose that the resistivity tensor is a function of any parameters, which change sign after time inversion.  
Usually one considers magnetic field, but also dependence on any other vector $\vec X$, which changes sign after reversal of the time, can be appropriate. Thus   $\rho_{yx} = \rho_{yx} (\vec{B},\vec{X})$, where $\vec{X}$ can be internal magnetisation $\vec{M}$ for example.\\ 

To simplify this picture we will discuss following  cases:\\
{\it{i}}) $B = 0$, $X = 0$ , $\rho_{xx} (\vec{0},\vec {0}) \neq \rho_{yy} (\vec{0},\vec {0}) $ (the anisotropy of the resistivity).
One can observe a transversal electric field $E_{\perp}= [(\rho_{xx}- \rho_{yy})cos\theta sin\theta]j$.
This transversal field is observed in zero magnetic field - "the zero field effect". \\
{\it{ii}}) $B \neq 0$, $X = 0$, $\rho_{xx} (\vec{B}, \vec{0}) =  \rho_{yy} (\vec{B}, \vec{0})$ (the isotropic resistivity in magnetic field). 
One can observe a transversal electric field $E_{\perp}= \rho_{yx}j$, which  is due to antisymmetrical part of the resistivity tensor. This field is odd in $\vec{B}$ as a consequence of the Onsager's relations and represents the Hall effect.\\
{\it{iii}}) $B \neq 0$, $X = 0$,  $\rho_{xx} (\vec{B},\vec {0}) \neq \rho_{yy} (\vec{B},\vec {0})$ (the anisotropy of the resistivity in magnetic field). 
One can observe the Hall effect together with a transversal electric field due to resistivity anisotropy. This transversal field is even in $\vec{B}$ and represents conventional "even effect" (or the transverse even effect).\\
{\it{iv}}) $B = 0, X \neq 0$, $\rho_{xx} (\vec{0}, \vec{X}) =  \rho_{yy} (\vec{0},\vec {X})$ (the isotropic zero-magnetic field resistivity under influence of the additional anisotropy determined by vector $\vec{X}$)).   
One can see a transversal electric field, which is due to antisymmetric part of the resistivity tensor. This field does not originate from crystallographic anisotropy (or anisotropy due to planar defect)  but from additional anisotropy due to $\vec{X}$. This transversal field can be observed in zero magnetic field, thus this effect is also "zero field effect". This field must be odd in $\vec{X}$ as a consequence of the Onsager's relations.\\  
{\it{v}}) $B  \neq 0, X \neq 0$ , $\rho_{xx} (\vec{B}, \vec{X}) =  \rho_{yy} (\vec{B},\vec {X})$ (the isotropic resistivity  in magnetic field and under additional  anisotropy determined by vector $\vec{X}$).
One can observe the Hall effect together with a transversal electric field, which can be even in magnetic field $\vec{B}$. This effect is thus  also an "even effect". For the antisymmetric part of the tensor we can say that  $\rho_{yx} (\vec{B},\vec {X}) = - \rho_{yx} (-\vec{B}, -\vec{X})$, but in general case  $\rho_{yx} (\vec{B},\vec{ X})  \neq - \rho_{yx} (-\vec{B}, \vec{X})$ and $\rho_{yx}$  can thus be separated into two parts, which are  odd and even function of the $\vec B$, respectively. \\

Finally we prove, that if  $X = 0$, the Hall resistivity cannot be anisotropic. For odd part of the nondiagonal element the next formula must be valid\\\\
$\rho_{xy}^{(odd)} (\vec{B}) =  -\rho_{xy}^{(odd)} (-\vec{B})$\\\\
and also Onsager relation holds in the form\\\\ 
$\rho_{xy}^{(odd)} (-\vec{B}) =  \rho_{yx}^{(odd)} (\vec{B})$,\\\\
because $X = 0$. Hence the next relation follows\\\\
$\rho_{xy}^{(odd)} (\vec{B}) =  -\rho_{yx}^{(odd)} (\vec{B})$,\\\\ 
which implies isotropic Hall resistivity. 

From the above discussed  cases the next conclusions follow. An anisotropy, which is the crystallographic structure anisotropy or which is caused by  some planar defects in the role of a guiding pinning canal, can evoke existence of the even effect and zero field effect as well. These anisotropies can be described by a vector, but  it does not change sign after time inversion. Thus these anisotropies cannot explain invalidity of the reciprocity theorem and also the anisotropic  Hall resistivity (including the odd longitudinal resistivity observed elsewhere \cite{Prod}). We must moreover assume the existence of an additional variable, which changes sign after time inversion. This idea  can explain not only the anisotropic Hall resistance and existence of the even effect, but also the zero field effect - observation of the transversal voltage in  high temperature superconductors in zero magnetic field. However, the origin of this quantity is unknown. One can speculate, that it can be for example the spontaneous magnetisation observed in YBaCuO material \cite{Car} or directly order parameter, which breaks time reversal symmetry \cite{Hor}. 
\section{Conclusion}
We have detected transversal voltages which are even and odd functions of the magnetic induction, respectively, in highly c-axis oriented YBaCuO superconductor. For low magnetic field the even voltage is higher than the odd Hall voltage. Temperature dependence of the even transversal voltage is similar to the non-zero transverse voltage measured in zero magnetic field. For higher field (in our experiment above about 0.2 T) even effect gradually vanishes. For high magnetic field (5 T) we have observed only the Hall effect, which shows anisotropy just below the critical temperature. The observed even effect can be explained in terms of  guided motion model. However, the  detected anisotropy of the Hall resistivity as well as reciprocity theorem breaking lead to an  assumption of existence some characteristic, which is independent on the external magnetic field and which changes sign after time inversion. It could be  for example spontaneous magnetisation observed in YBaCuO materials.

\newpage
\section*{Figure captions}

Fig. 1.: Temperature dependence of resistivities:   $ R^{\mathrm{(III)}}_{ABCD}$  (full circle)  and $ R^{\mathrm{(III)}}_{DABC}$  (square)  for $(a)$ $B=0$ T; resistivities  $R^{\mathrm{(III)},{even}}_{ABCD}$  (full circle), $R^{\mathrm{(III)},{even}}_{DABC}$  (square), $R^{\mathrm{(III)},{odd}}_{ABCD}$  (full line), $R^{\mathrm{(III)},{odd}}_{ABCD}$  (dashed line) for  (b) $B=0.05$ T, (c) $ B= 0.2$ T, (d) $B=0.5$ T, (e) $B= 5$ T. \\\\
Fig. 2.: Magnetic field dependence of the resistivities  $R^{\mathrm{(III)}}_{ABCD}$  (full circle), $R^{\mathrm{(III)},{even}}_{DABC}$  (dashed line), $R^{\mathrm{(III)},{even}}_{ABCD}$ (dotted line) for $(a)$ $T= 89$ K and $(b)$ $T= 89.2$ K.\\\\
Fig. 3.: Temperature dependence of $A$ for different magnetic fields $(a)$  $B= 0$ T  (dotted line), $B=+0.05$ T  (full circle), $B= -0.05$ T  (open circle); $(b)$ $B= + 5$ T  (full circle), $B= -5 $T (open circle).
    
\end {document}